\documentstyle{mn}

\def\ie{{i.e. }}

\def\vs{{\vskip 1.0ex}}
\newcommand{\dgr}{{^\circ}}
\newcommand{\gapprox}{$_ >\atop{^\sim}$}

\input psfig

\title[A survey for high-redshift radio-loud quasars]{
A survey for high-redshift radio-loud quasars: optical spectroscopy of
S$>$0.2 Jy, flat-spectrum radio sources}

\author[I. M. Hook et al.]
{\parbox[]{6.in} 
{I. M. Hook$^1$, R. G. McMahon$^2$, M. J. Irwin$^3$ and C. Hazard$^{2,4}$} \\
       $^1$ U.C. Berkeley Astronomy Dept., Berkeley CA94720, USA, e-mail imh@bigz.berkeley.edu\\
       $^2$ Institute of Astronomy, Madingley Road, Cambridge CB3 0HA, UK \\
       $^3$ Royal Greenwich Observatory, Madingley Road, Cambridge, CB3 OEZ, UK \\
       $^4$ University of Pittsburgh, Pittsburgh, PA 15260, USA}

\begin{document}
\maketitle

\begin{abstract} We present optical spectroscopic data for
a complete sample of 161 $\rm S_{5GHz}\ge 0.2$Jy, flat-spectrum radio
sources. The sources were observed as part of a survey for high
redshift, radio-loud quasars, and were selected for spectroscopic
follow-up based on criteria of red optical colour and unresolved
optical counterpart, as measured from APM scans of POSS-I plates.
13 objects from the spectroscopic sample were found to be
radio-loud quasars with $z>3$, of which two were previously known.  We
give positions, E (red) magnitudes, $\rm O-E$ colours, 5GHz radio
fluxes, radio spectral indices, optical spectra and redshifts where
possible for the spectroscopic sample. We also give finding charts for
the $z>3$ QSOs. The highest redshift object found is a QSO with
$z=4.30$ (GB1508+5714, the subject of an earlier Letter). The sample
also contains a $z=3.05$ QSO, GB1759+7539, which is optically very
luminous (E=16.1). In addition, spectra are given for 18 $\rm S_{5GHz}\ge
0.2$Jy, flat-spectrum radio sources that do not form part of the
complete sample.

\end{abstract} 
\maketitle

\section{Introduction}

The first quasar to be identified as such was the strong radio source
3C 273, with a redshift of $0.158$ (Hazard, Mackey \& Shimmins 1963,
Schmidt 1963).  Since then, radio-selected samples have played an
important part in the discovery of quasars and in the study of their
evolution.  $V/V_{\rm max}$ analysis of a complete sample of
radio-selected quasars first led to the conclusion that the quasar
population is evolving rapidly (Schmidt 1968). Using quasars from the
3CR catalogue, Schmidt found that at constant luminosity, the space
density at $z=1$ is about 150 and 80 times the local density, for
$q_0=0$ and 1 respectively.

Over the past years, much observational effort has gone into the
construction of complete samples of quasars in order to determine the
luminosity function and its evolution with redshift.  
The status of the major quasar surveys and selection techniques have
been reviewed by Hartwick \& Schade (1990) and in two conference
proceedings (Osmer et al. 1988; Crampton 1991).  At present the
quasar population is believed to undergo strong luminosity evolution
up to redshifts of $\sim 2$, and this evolution slows down between
redshifts 2 and 3.  Prior to the work described in this paper, the
only large complete samples of QSOs with redshifts greater than 3
have been produced from optical surveys (Warren, Hewett \& Osmer 1991;
Irwin, McMahon \& Hazard 1991; Schneider, Schmidt \& Gunn 1994). The
form of the evolution at these redshifts is less clear-cut than at
lower redshifts, and it appears that a combination of luminosity
evolution and density evolution is taking place. The space density of
the less optically luminous quasars ($M_B\sim -26$) appears to decline
rapidly at $z>3$ (Schmidt, Schneider \& Gunn 1991, Warren, Hewett \&
Osmer 1994), whereas no evidence for a decline was found for the
higher luminosity optically selected quasars ($M_B<-28$) (Hazard,
McMahon \& Sargent 1986; Miller, Mitchell \& Boyle 1990; Irwin et
al. 1991).

Evidence for a decline in space density at high redshifts suggests
that the epoch of galaxy formation may have been found (with the
caveat that quasars probably represent the extreme of the matter
distribution and are not necessarily tracers of the galaxy
population).  There is, however, much controversy over these results
owing to the complex, redshift-dependent selection effects which
affect optical surveys for high-redshift quasars.

Radio selection methods are less prone to redshift-dependent bias, and
the resulting quasar samples can be used to derive the luminosity
function independently of optical surveys.  Radio fluxes are
unaffected by any obscuration arising from dust in foreground
galaxies. Also, the spectral energy distribution of quasars at radio
wavelengths is smooth and is not complicated by either emission lines
or redshift-dependent intervening absorption so that the effects of
{\it k}-corrections are easier to model. Moreover, since the mean
radio spectral index of core-dominated, flat-spectrum radio sources is
about 0.0, a higher fraction will lie at high redshift compared with
samples selected at optical or X-ray wavelengths where spectral
indices are typically in the range $-$0.5 to $-$1.0.

In order to find sufficient quasars for a statistical analysis of
their space density, it is necessary to survey a large area or go to a
faint flux limit.  Previous work using radio samples has concentrated
on the optical identification of optically faint radio galaxies from
bright, large-area radio surveys, e.g. 3C and 4C, or weak radio
sources from deep surveys over small areas.  These samples were
usually selected at low frequency (e.g. 178MHz), which biases their
content to a high fraction of steep-spectrum sources, usually
identified with galaxies.  Surveys at higher frequency contain a
higher fraction of flat-spectrum, core-dominated sources, usually
identified with quasars.

Optical identification of sources from large-area radio surveys has
become more feasible since the advent of measuring machines for
determining accurate optical positions and magnitudes from
photographic plates.  In order unambiguously to identify high-redshift
quasars, which have similar colours to those of the stellar sequence,
accurate radio positions are needed. The use of the Very Large Array
(VLA) to measure positions of sources detected using other instruments
has made available large samples with positions accurate to better
than 1 arcsec.  The use of Schmidt telescopes and radio selection
techniques in quasar surveys is described by Hazard (1986).

Recent work in the identification of complete samples of radio sources
and their use in luminosity function calculations includes that of
Peacock (1985) and Dunlop \& Peacock (1990), in which redshifts have
been obtained for four radio samples selected at 2.7GHz. The faintest
of these samples reaches a flux limit of 0.1Jy and the combined
samples contain a total of nearly 200 radio-loud quasars.  However,
owing to the selection frequency and the small area covered at low
flux limits, these samples contain very few high-redshift
quasars. Only two quasars have spectroscopically confirmed redshifts
above $z=3$ in the combined flat- and steep-spectrum samples of
Dunlop \& Peacock (1990).  New data at high redshifts are therefore
needed to derive the luminosity function of radio-loud quasars with
$z>3$.

We have developed a technique to find radio-loud quasars with
$z>3$. The technique involves optical identification of a large number
of radio sources, followed by selection of the optically red,
unresolved objects. The first results were published in a recent
Letter (Hook et al. 1995). The current paper is concerned with the
results of spectroscopy of a complete $S_{\rm 5GHz}\ge 0.2$Jy
flat-spectrum sample which is briefly described in the following
section.  The radio data and optical identification procedure will be
described in detail in a future paper. The optical identification of
the radio sample and selection of the spectroscopic sample are
described in Section 3 of this paper. The observations and data
reduction are described in Sections 4 and 5 respectively, and the
results are presented in Section 6.  Finally, tables of redshifts,
optical and radio data for the spectroscopic sample and the extra
objects observed are given, along with the spectra.

\section{The Radio Data}

Patnaik et al. (1992a) have analysed 5-GHz radio maps from the $S$
\gapprox\ 25mJy Green Bank (GB) survey (Condon, Broderick \& Seielstad
1989) in order to produce a sample of flat-spectrum sources brighter
than 0.2Jy. Spectral indices were calculated from the 5-GHz flux
densities and the 1.4GHz flux densities from Condon \& Broderick
(1985,1986).

About 800 flat-spectrum ($\alpha_{1.4}^5 \ge -0.5$,
$S\propto\nu^{\alpha}$, where $\alpha_{1.4}^5$ is the spectral index
between 1.4 and 5-GHz) sources in the declination range
$35\dgr\le\delta({\rm B1950})\le 75\dgr$ and with $|b|\ge 2\dgr.5$
were observed with the VLA (Patnaik, et al. 1992a) giving positions
with rms accuracy of $\sim$ 12 milliarcsec.  A further $\sim800$
flat-spectrum sources in the ranges $20\dgr\le\delta({\rm B1950})\le
35\dgr$ and $75\dgr\le\delta(B1950)\le90\dgr$ were also observed.
Those sources with $\delta >75\dgr$ originated from the S5 survey
(K\"uhr et al. 1981) and spectral indices for these sources were taken
from K\"uhr et al. (1981).  The radio sample therefore contains $\sim
1600$ sources observed with the VLA, and covers an area of 4.1sr
(14000 $\rm deg^2$).

Although the 5-GHz flux densities determined by Patnaik et al. were
used to define the original radio sample, they have been found to be
systematically higher by 10-15 per cent than the flux densities
determined by Gregory \& Condon (1991) from the same maps.  Gregory \&
Condon have published a comprehensive description of the methods used
to determine their flux densities, including a comparison with the
Becker, White \& Edwards (1991) flux determinations from the same
maps.  The flux densities were also compared with more accurate
measurements at 5-GHz (Kulkarni, Mantovani \& Pauliny-Toth 1990) of
steep-spectrum sources (i.e. predominantly lobe-dominated and
non-varying) from the 408-MHz B3 survey.  In this paper we have used
5-GHz flux densities from Gregory
\& Condon (1991) and 1.4-GHz flux densities from White
\& Becker (1992) (e.g. Table~\ref{knowntab}).  For sources with $\delta
> 75\dgr$ we have used 5-GHz flux densities from the S5 catalogue (K\"uhr et
al. 1981), and, for sources with $\delta > 82\dgr$, above the
declination limit of the 1.4-GHz survey, spectral indices between
2.7 and 5-GHz from K\"uhr et al. (1981) were used.  A full
comparison of the flux densities will be made in a future paper.  We
have calculated the effective flux limit of the Patnaik et al. sample
to be $\sim 175$mJy. Since Patnaik et al. used high values for the
5-GHz flux, the spectral indices of the sources, $\alpha^5_{1.4}$, were
estimated to be slightly too flat (i.e. $\alpha$ more positive if
$S\propto \nu ^{\alpha}$) by $\sim0.1$.  The effective spectral index
limit of the sample is therefore about $-0.6$. Occasional steeper
(more negative) values of $\alpha$ will be seen in
Tables~\ref{knowntab},
\ref{comptab} and \ref{extratab} since we have used different 1.4-GHz flux densities
from those used by Patnaik et al. (1992a) to define the radio sample.

This sample has an advantage over other radio samples for determining
the quasar luminosity function in that the contribution from strong
lensing will be known explicitly, since all the objects have been
observed at high resolution (0.24 arcsec FWHM) using the VLA. Follow
up observations to confirm the lenses are underway. At present the
known lenses contained in the radio sample are 1422+2309 ($z=3.62$,
Patnaik et al. 1992b), 1938+666 (Patnaik \& Narasimha 1993) and
0218+357 (Patnaik et al. 1993).

\section{Optical identification and selection of the Spectroscopic sample}
Optical identification of the radio sources was carried out using the
Automated Plate Measurement (APM) Facility at Cambridge, U.K.  The
process involves the digitisation and internal calibration (Bunclark
\& Irwin 1983)  of Palomar Sky Survey (POSS-I) photographic plates
in two passbands, E (red) and O (blue).  Plates were scanned with
$\sim0.5$-arcsec pixels, and the coordinate systems of the two plates
covering each field were matched. E and O magnitudes were measured for
each detected object, and the images were classified as galaxies,
`stars' (\ie\ unresolved images) or noise. Various combinations of
image parameters are used for this, e.g. moments, ellipticity, etc.
The spread in the distribution of each parameter with respect to the
stellar locus is found and is used to weight that parameter when they
are combined to form a final classification parameter.  The mean and
spread of the distribution of the combined parameter are found as a
function of magnitude, and a statistical correction calculated to
convert these to Gaussian distributions with zero mean, rms=1. This is
done only for the `stellar' objects. In a second pass the correction
is applied to all the images and then the number of standard
deviations from a stellar profile, $N\sigma_c$, is recorded.  For
images about 2 magnitudes above the plate limit, the classification is
about 90 per cent accurate.

Optical identifications of the radio sources were made on the basis of
positional coincidence alone. The rms difference between optical and
radio positions, $\sigma_r$, was determined by fitting a Rayleigh
distribution function to the distribution of $\Delta r$, and was found
to be 0.7 arcsec. The criterion for positional coincidence was chosen
to be $\Delta r \le 3.0$ arcsec, i.e. $\sim 4.3\sigma_r$. At larger
separations the number of background red, stellar sources begins to
dominate the number of real red, stellar identifications.  Assuming
Gaussian statistics, this results in a negligible fraction ($<0.01$
per cent) of correct identifications being lost from the sample. In
practice the fraction is probably somewhat higher than this owing to
non-Gaussian errors which cause the distribution of $\Delta r$-values
for true identifications to have a longer tail.  The background source
density of red, stellar objects was estimated from the annulus 10 to
15 arcsec around each source and was found to be 0.002 $\rm
arcsec^{-2}$. This gives a false identification rate in a 3.0-arcsec
radius of 0.057.

The zero-point for photometry in each field was calculated by assuming
that the E plate has a limiting magnitude of 20.0 and by assuming a
universal, magnitude-independent position in the colour-magnitude
plane for the stellar locus, as described by McMahon (1991).  The
zero-point of the magnitude system has an rms uncertainty of 0.25mag.
A description of the plate material and the steps in the plate
measurement and candidate selection, along with a full list of the
optical identifications, magnitudes and image classification, will be
given in a future paper.

The sample of candidate high-redshift quasars was selected using the
APM image classification, O$-$E colour and positional coincidence of
the radio source with an optical counterpart. The basis for the colour
selection method is that the $\rm O-E$ colours of QSOs with $z>2$
become redder rapidly with redshift, owing to absorption by intervening
Ly$\alpha$ (see fig. 1 in Hook et al. 1995). To define a sample with
a high level of completeness at $z>3$ we chose a limit of $\rm O-E\ge
1$.

$N\sigma_c\le 3.0$ was chosen as the classification criterion based on
the distribution of $N\sigma_c$ for blue ($\rm O-E < 1.0$) objects
within 3.0 arcsec of each radio source, most of which will be QSOs
(with $z<3$). 92 per cent of these blue identifications have $N\sigma
\le 3.0$. A limit of $\rm E\ge 11.0$mag was also imposed on the
spectroscopic sample since $z>3$ QSOs are not expected to be brighter
than this.

In summary, the selection criteria were (i) positional coincidence,
$\Delta r\le 3$ arcsec, (ii) red optical colour $\rm O-E\ge 1.0$, (iii)
unresolved image on the E plate, $N\sigma_c\le 3.0$, (iv) $\rm E \ge 11.0$
mag. In addition, the optical counterpart had to be bright enough to
appear on the E plate, \ie $\rm E\le 20.0$. The counterpart was not
required to be detected on the O plate.

Of the initial radio sample, $\sim 940$ sources lie in the region
covered by the APM scans (i.e. avoiding the galactic plane, $|b|\ge
20-30\dgr$), an area of 2.2sr ($\rm 7300 deg^2$). 738 of these have
optical identifications within 3 arcsec.  Of these, 161 satisfy the
criteria above. This sample of 161 objects will be referred to as the
`complete sample'.

\begin{figure*}
\centering
\mbox{\psfig{figure=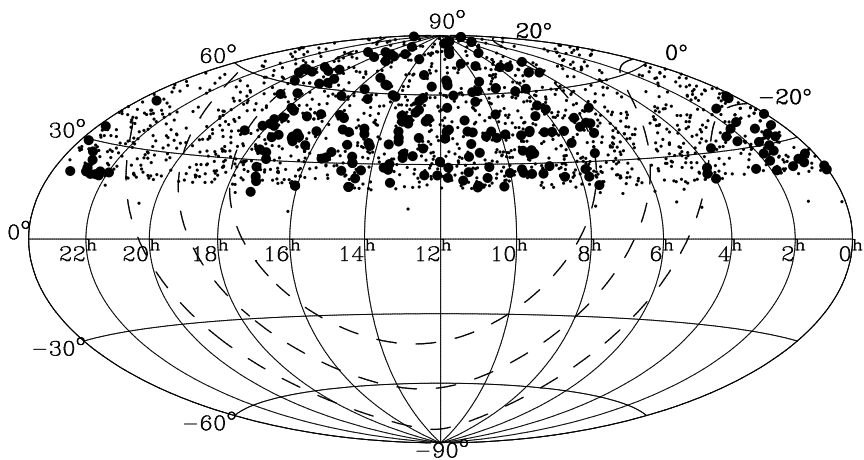,width=4.5in,bbllx=70pt,bblly=274pt,bburx=537pt,bbury=524pt}}
\caption[]{Aitoff projection showing the sky distribution of the parent radio 
sample (points) and identifications satisfying the selection criteria,
i.e. the complete spectroscopic sample (filled circles). Dashed lines show the
galactic equator and galactic lines of latitude $|b|=20\dgr$.}
\label{aitsky}
\end{figure*}

\begin{figure}
\centerline{
\psfig{figure=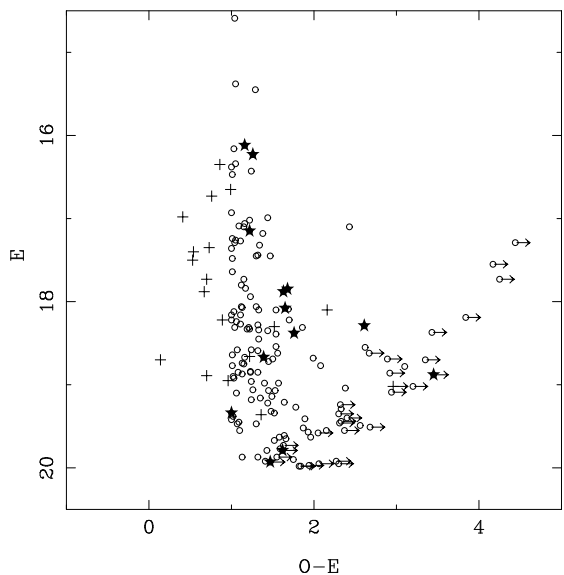,height=3.0in,bbllx=39pt,bblly=123pt,bburx=525pt,bbury=615pt}
}
\caption[]{Colour-magnitude diagram for the optical identifications of
the 0.2Jy sample. The open circles represent objects satisfying the
selection criteria for the spectroscopic sample. Crosses show
additional objects for which we have obtained spectra. Arrows show
limits on $\rm O-E$ colour for objects that were not detected on the O
plate. The $z>3$ QSOs are shown by star symbols.}
\label{jbcolmag}
\end{figure}

Fig.~\ref{aitsky} shows the distribution on the sky of the parent
radio sample and the complete spectroscopic sample, and a
colour-magnitude diagram showing the complete sample is given in
Fig.~\ref{jbcolmag}. Prior to the start of this project in 1992, 38 of
the complete sample had published redshifts, including two with
$z>3$. These two are 0636+6801 with $z=3.18$ (an S4 source: K\"uhr
1977, 1980) and 0642+4454 with $z=3.41$ (OH471: Gearhart et al. 1972,
Carswell \& Strittmatter 1973).  This left 123 objects requiring
spectroscopy.  The previously known objects were not re-observed
(except one, GB0916+8625).  Their properties are summarized in
Table~\ref{knowntab}. There are no known radio-loud, $z>3$ quasars
within the parent radio sample that do not fulfil the colour criterion
$\rm O-E>1.0$.

\begin{table*}

\caption{Objects in the complete sample that had published redshifts
prior to the start of the project in 1992. These objects were taken
from Hewitt \& Burbidge (1987, 1989, `HB') and from V\'eron-Cetty \& V\'eron
(1991, `V91'). Redshifts and object classifications are from NED$\dag$
(references given there) except where stated.}


\begin{center}
\begin{tabular}{r@{\hspace{0.2cm}}     r@{\hspace{0.2cm}}rr@{\hspace{0.2cm}}r@     {\hspace{0.2cm}}rccrrrrrl}
\hline
\multicolumn{6}{c}{Optical Position}&       \multicolumn{1}{c}{$z$} &       $N\sigma_c$    &\multicolumn{1}{c}{E}       & O$-$E &$\Delta r$      &\multicolumn{1}{c}{$S_{5}$}       & \multicolumn{1}{c}      {$\alpha^{5}_{1.4}$}      & Comments \cr
\multicolumn{2}{c}{$\alpha$}&   \multicolumn{2}{c}{1950}&   \multicolumn{2}{c}{$\delta$}& &&mag &mag&     \multicolumn{1}{c}{$''$} &mJy& & \cr\hline
 01&09&23.58& 22&28&44.3& $-$&0.12&14.59&$  1.04$& 0.55&   260&$   -0.11\phantom{^*}$& BL Lac \cr 
 01&49&31.70& 21&52&19.9&1.32&0.55&18.06&$  1.12$& 0.99&  1060&$   -0.21\phantom{^*}$& QSO \cr 
 02&01&56.24& 36&34&57.0&2.91&0.68&17.45&$  1.30$& 1.06&   349&$   -0.41\phantom{^*}$& QSO \cr 
 04&09&44.67& 22&57&27.5&1.21&0.44&18.10&$  1.54$& 0.03&  1000&$   -0.20\phantom{^*}$& QSO \cr 
 04&54&57.15& 84&27&52.7&0.11&0.77&17.10&$  1.15$& 0.38&  1398&$    0.38^*          $& QSO \cr 
 06&36&47.38& 68&01&25.9&3.18&1.95&16.23&$  1.26$& 1.83&   499&$    1.06\phantom{^*}$& QSO \cr 
 06&42&53.03& 44&54&30.5&3.41&0.67&17.88&$  1.63$& 0.37&  1191&$    0.55\phantom{^*}$& QSO; $z$ is from V91, $z$ = 3.396 in NED \cr 
 07&10& 3.31& 43&54&26.3&0.52&0.72&19.73&$> 1.63$& 0.27&  1629&$   -0.09\phantom{^*}$& QSO \cr 
 07&16& 4.43& 33&12&42.7& $-$&1.24&18.06&$  1.30$& 1.07&   364&$    0.44\phantom{^*}$& listed in V91, no optical ID in NED \cr 
 07&16&12.83& 71&26&15.0& $-$&0.05&14.42&$  1.03$& 0.88&   788&$   -0.02\phantom{^*}$& BL Lac \cr 
 07&45&35.75& 24&07&55.3&0.41&1.97&18.31&$  1.04$& 0.45&  1196&$    0.15\phantom{^*}$& QSO \cr 
 08&14&51.64& 42&32& 7.0&0.26&0.28&18.85&$  1.08$& 0.90&  1891&$    0.20\phantom{^*}$& BL Lac \cr 
 08&27&54.37& 24&21& 7.1&2.05&1.33&16.16&$  1.03$& 0.75&   672&$   -0.03\phantom{^*}$& QSO; HB give $z=0.94$\cr 
 08&28&47.96& 49&23&32.2&0.55&1.11&18.34&$  1.33$& 0.90&   363&$   -0.80\phantom{^*}$& BL Lac \cr 
 08&49& 4.49& 28&45&18.3&1.27&0.53&19.34&$  1.52$& 0.92&   422&$    0.19\phantom{^*}$& QSO \cr 
 09&16&11.95& 86&25&16.9&    &0.56&18.07&$  1.13$& 1.23&   253&$              0.21^*$& BL Lac; also observed in this work   \cr                                      
 09&17&40.24& 62&28&37.9&1.45&1.01&18.16&$  1.11$& 1.00&  1322&$    0.06\phantom{^*}$& QSO \cr 
 09&54&57.72& 65&48&15.1&0.37&0.19&15.45&$  1.29$& 0.94&  1417&$    0.61\phantom{^*}$& BL Lac \cr 
 10&30& 7.85& 41&31&34.1&1.12&0.64&17.36&$  1.00$& 0.70&   485&$   -0.37\phantom{^*}$& QSO \cr 
 10&48&10.41& 34&46& 5.6&2.52&1.43&18.98&$  1.40$& 0.36&   298&$   -0.46\phantom{^*}$& QSO \cr 
 10&53&35.84& 81&30&36.3&0.71&0.09&18.09&$  1.69$& 1.08&   770&$    0.64\phantom{^*}$& Galaxy \cr 
 11&38&40.07& 65&52&59.9&0.81&1.28&17.73&$> 4.25$& 0.84&   184&$   -0.69\phantom{^*}$& Galaxy \cr 
 12&31&50.22& 48&10&22.6&0.38&0.91&16.47&$  1.01$& 0.31&   278&$   -0.23\phantom{^*}$& Listed in V91, no ID in NED \cr 
 12&54& 4.89& 57&08&37.3&0.04&1.84&11.08&$  1.15$& 1.46&   419&$    0.29\phantom{^*}$& Galaxy \cr 
 13&08& 7.54& 32&36&39.4&1.00&2.92&18.59&$  1.32$& 1.06&  1131&$   -0.28\phantom{^*}$&BL Lac \cr 
 14&13&22.73& 37&20&15.5&2.36&0.08&17.29&$  1.04$& 0.77&   383&$    0.04\phantom{^*}$& QSO \cr 
 14&17&43.05& 38&35&30.6&1.82&0.04&19.40&$> 2.40$& 1.83&   871&$    0.16\phantom{^*}$& QSO \cr 
 14&24&44.17& 24&01&25.0& $-$&0.33&15.38&$  1.05$& 1.16&   316&$   -0.24\phantom{^*}$& BL Lac \cr 
 14&42&52.98& 63&45& 3.8&1.38&0.33&17.27&$  1.11$& 0.58&   456&$   -0.32\phantom{^*}$& Galaxy \cr 
 15&20&56.25& 34&24&46.4&1.31&0.79&18.22&$  1.00$& 0.70&   147&$   -0.34\phantom{^*}$& QSO \cr 
 16&38&48.12& 39&52&30.3&1.66&0.39&17.24&$  1.01$& 0.70&  1285&$    0.53\phantom{^*}$& QSO \cr 
 17&22&23.36& 33&05&42.9&1.87&0.31&18.85&$  1.32$& 0.75&   435&$    0.23\phantom{^*}$& QSO \cr 
 17&32&40.46& 38&59&46.0&0.97&0.01&18.87&$  1.13$& 1.09&   561&$   -0.26\phantom{^*}$& BL Lac \cr 
 17&38&12.61& 49&56&35.5&1.55&0.24&17.48&$  1.01$& 0.86&   478&$   -0.13\phantom{^*}$& QSO \cr 
 18&03&39.27& 78&27&54.0&0.68&1.61&13.93&$  1.50$& 0.31&  2633&$    0.27\phantom{^*}$& BL Lac \cr 
 18&23&14.94& 56&49&18.0&0.66&1.19&17.02&$  1.22$& 0.22&  1135&$   -0.21\phantom{^*}$& QSO \cr 
 22&17&15.98& 21&26& 6.4&1.52&0.48&18.62&$> 2.67$& 1.20&   248&$   -0.12\phantom{^*}$& QSO \cr 
 22&51&44.32& 24&29&23.6&2.33&0.27&17.26&$  1.05$& 1.30&   937&$   -0.62\phantom{^*}$&QSO \cr 
\hline 
\end{tabular} 
\end{center} 
\flushleft

\noindent{\bf Notes:} \newline \noindent{\bf Optical data:} APM
measurements of optical positions, E magnitudes, O$-$E colours and the
classification parameter $N\sigma_c$ are described in McMahon et al.
(in preparation).\newline
\noindent{\bf Radio data:} 5GHz fluxes $S_5$ from Gregory \& Condon
(1991) are given for sources with $\delta < 75\dgr$. For higher
declination sources, 5GHz fluxes are taken from the S5 catalog (K\"uhr
et al. 1981).  For $\delta < 82\dgr$, the spectral indices
$\alpha_{1.4}^5$ ($S\propto\nu^{\alpha}$) were calculated using these
5GHz fluxes and 1.4GHz fluxes from White \& Becker (1992). Limits are
given for objects below the flux limit of the 1.4GHz catalog,
100mJy. For $\delta > 82\dgr$, above the declination limit of the
1.4GHz catalog, we give the 2.7GHz to 5GHz spectral indices from K\"uhr
et al. (1981), and mark these objects with a `$*$' in the $\alpha_{1.4}^5$
column.

\noindent $\dag$ The NASA/IPAC Extragalactic Database (NED) is
operated by the Jet Propulsion Laboratory, California Institute of
Technology, under contract with the National Aeronautics and Space
Administration.

\label{knowntab}

\end{table*}

\section{Optical Spectroscopy}
The aim of the spectroscopic phase of this project was to obtain
classification (high-redshift QSO or not) for the complete
spectroscopic sample defined above.  This does not require redshifts
to be obtained for all the objects: for example, many `featureless'
spectra could be rejected as possible $z>3$ quasars by the lack of
Ly$\alpha$ forest absorption.  The strategy was therefore to take
short exposures at low dispersion to obtain spectra of sufficient
quality for the nature of the object and, where possible, the redshift
to be determined.  In a few cases, where the objects turned out to be
particularly interesting, longer exposures were taken (GB1508+5714 and
GB1745+6227).  In poor weather, spectra were taken of optically
brighter identifications which do not necessarily satisfy all the
selection criteria described in the previous section. Some of these
are slightly resolved on the basis of the APM structure parameter,
$N\sigma_c$, and some are bluer than the limit $\rm O-E = 1$.

\subsection{Observational set-up}
Most of the spectroscopy was carried out in 1992 October and 1993
April on the 4.2-m William Herschel telescope (WHT), La Palma.  The
observations were taken using the Durham-RGO Faint Object Spectrograph
(FOS-II) or the ISIS double spectrograph.  Typical exposure times were
300-600s.

\subsubsection{Faint Object Spectrograph} FOS 
is a fixed-format, low dispersion spectrograph of high throughput,
specifically designed for observing faint objects. FOS-II is used at
the $f/11$ Cassegrain focus of the WHT. A $385\times 578$
GEC CCD, coated to improve UV response, was used as a detector.
Dispersion is provided by a transmission grating and a
cross-dispersing prism which together give a two-order format covering
the wavelength range 4000 to 10500\AA.  For this project useful data
were obtained only in first order (8.7\AA $\rm pixel^{-1}$ on the WHT)
over the wavelength range 4800--9500\AA.  FOS-II is described in more
detail by Allington-Smith et al. (1989).

\subsubsection{ISIS double spectrograph}
ISIS is a double spectrograph with arms optimized for blue and red light
which was used at the $f/11$ Cassegrain focus of the WHT.  A range of
intermediate dispersions are possible from 120 to 8\AA$\rm mm^{-1}$. 
For this project the lowest dispersion was required, and a grating with
158 lines/$mm$ and a dichroic at $\sim$5400\AA\ were used.  The gratings
were arranged so that the blue part of the spectrum was centred on
4600\AA\ and covered a range of 2950\AA\ while the red was centred on
7000\AA\ and covered a range of 3380\AA. 

On the red arm an EEV $1242\times 1152$ 
CCD with $\rm 22.5\mu m$ pixels was used as detector. On the blue arm
a thinned Tektronix CCD with
$1024\times 1024$ ($\rm 24\mu m$) pixels was used.

\subsection{Observations} A total of six nights were allocated to this
project in 1992 October and 1993 April.  Prior to that, some
observations were carried out interspersed with other
programmes. Table~\ref{obsdates} shows the instrument used during each
run, and the number of objects observed from the complete sample.  The
observations on June 28/29 and August 17-19 were carried out in the
context of the PATT service observation programme.

\begin{table}
\centering
\caption[]{Dates of observations, instruments used and number of objects
observed from the complete sample.}
\vs
\vs
\begin{tabular}{|l@{\hspace{0.2cm}}l@{\hspace{0.2cm}}l@{\hspace{0.2cm}}lr}\hline
\multicolumn{3}{|c|}{UT date}& Instrument & \multicolumn{1}{c}{No. Observed}\cr \hline
$20 - 25$,  27 &Apr&1992& FOS-II/ISIS &22 \cr 
$02 - 04$      &Oct&1992& FOS-II/ISIS &21 \cr 
$09 - 12$, 17  &Apr&1993& FOS-II/ISIS &75 \cr 
28             &Jun&1993& ISIS        &1  \cr 
$17-18$        &Aug&1993& ISIS        &1  \cr 
$9-10$         &Jun&1994& FOS-II      &2  \cr 
\hline
\hline
\end{tabular}
\flushleft
\label{obsdates}
\end{table}

For all observations, the width of the slit was adjusted to be
compatible with the seeing at the time of observation ($1-2$ arcsec). All
observations were taken with the slit at the parallactic angle. Almost
all the objects were too faint to be easily visible on the acquisition
TV, so accurate offsetting from nearby stars was used to position the
target object in the slit.  Spectrophotometric standards were observed
to calibrate the spectra.

All ISIS observations were made with a long slit ($\sim 2$
arcmin). The FOS-II observations were made with a short slit
(projected length $\sim 20$ arcsec) to prevent the first- and
second-order spectra overlapping.  The CCD chips were windowed in the
spatial direction to reduce the readout time.

Further observations of GB1508+5714 were carried out on 1993 April
17. Three 900s observations were obtained with the red arm and two
1500s observations with the blue arm, with a 1.5 arcsec spectrographic
slit. Red and blue observations of the flux standard HD 84937 were
used to flux-calibrate the data.

GB1745+6227 was re-observed on 1992 August 1 (a service night). Seven
900-s observations were obtained with the red arm, and six 900-s
observations with the blue arm. Red and blue observations of the flux
standard BD +40$\dgr$ 4032 were used to obtain relative flux
calibration.

\section{Data Reduction}
The reduction of data was carried out using standard software from the
IRAF package.  The steps taken in the data reduction are summarized
below.

\noindent{\it (1) Overscan correction, bias correction and flat-fielding.} 
A linear ramp was fitted to the overscan region in the direction
perpendicular to the dispersion axis, and was subtracted from the 2D
frames.  For each observing run, a combined bias frame was subtracted
from the 2D frames.

Flat-field frames were taken on each run using a tungsten lamp to
illuminate the slit, and were used to correct for pixel-to-pixel
sensitivity variations.

\noindent{\it (2) Extraction of spectra.}
The object spectra were located and traced along the chip.  Typically
the ISIS spectra curved by less than a pixel from one end of the chip
to the other, whereas the FOS-II spectra shift by about 35 pixels, as
a result of the cross-dispersing prism.  The functional form of the
tracing function was specified to be a Legendre polynomial of order 4
for FOS-II and order 3 for ISIS.  The sky level was determined by
interpolating between specified regions on either side of the
spectrum, and was subtracted.  The extraction was variance-weighted
using the variances determined from the data and CCD characteristics.
 
\noindent{\it (3) Wavelength calibration and flux calibration.}
Cu-Ar+Cu-Ne arc lamp spectra were used to wavelength calibrate the
data.  Non-blended emission lines were identified, and a
pixel-to-wavelength calibration curve was found by fitting a
third-order polynomial to the calibration points. Typical rms
residuals from the fit were 0.3\AA\ for FOS-II and 0.2\AA\ for ISIS.
A few spectra were calibrated using emission and absorption features
in the sky spectra. Typical {\it rms} residuals derived using sky
spectra were 0.3\AA\ for ISIS, and 0.8-1\AA\ for FOS-II.

Observations of spectrophotometric standards were used to
flux-calibrate the data.  Usually only relative flux calibration over
the wavelength range of each spectrum was required. Some observations
were carried out through thin and patchy cloud which causes
wavelength-independent extinction. The relative flux calibration is
unaffected, however, and the resulting spectra can still be used for
spectral index determination.

The flux calibration procedure was checked by flux-calibrating the
standards and overlaying the calibration points.  It was found that
calibration was reliable over the wavelength ranges 5500-8600\AA\
(ISIS red), 3600-5500\AA\ (ISIS blue) and 4950-9500\AA\ (FOS-II).

The high-quality ISIS spectra that were taken of the highest redshift
objects were corrected for atmospheric absorption using observations
of featureless B stars.

\subsection{Measurement of redshifts}
Redshifts were estimated for objects that had obvious emission lines.
For quasars with redshifts above $\sim 2.2$ redshifts could be
measured using the broad emission lines CIV (rest wavelength 1549\AA)
and CIII] (1909\AA).  Ly$\alpha$ (1216\AA) was also used to identify
these objects (depending on the wavelength coverage of the spectrum)
but the redshift based on that line is an upper limit since the peak
of Ly$\alpha$ is shifted towards the red due to absorption by the
Ly$\alpha$ forest. In some objects the NV (1240\AA) and SiIV/OIV]
(1400\AA) features could also be used to derive the redshift.  For
objects with redshifts less than $\sim 0.8$, the [OIII](5007/4959\AA)
and H$\beta$(4861\AA) emission lines could be identified.
Intermediate redshift objects could be identified by the MgII
(2798\AA) and CIII] emission lines.

The observed wavelengths were measured from the centroids of the lines
(calculated in IRAF) unless the line profile was visibly affected by
sky absorption. When more than one clean line was visible, an average
redshift was taken, otherwise the redshift was based on the cleanest
(usually strongest) line.

\section{Results and Discussion}



\begin{figure*}
\centerline{
}
\label{jbspec}
\caption[]{Objects satisfying the selection criteria of the complete sample.}
\end{figure*}

\begin{figure*}
\centerline{
}
\caption[]{Objects which did not satisfy all the selection criteria. The properties of these objects are summarised in Table~\ref{extratab}.}
\label{jbothers}
\end{figure*}

\begin{table*} \caption{Objects in the complete sample which we
have observed with WHT. Spectra for these objects are given in
Figure~ref{jbspec}.  See Table~1 for a description of the column
headings. Redshifts are given for objects with obvious emission
features. A `*' indicates an object with an uncertain redshift,
e.g. when only one line is visible.}
\begin{center}
\begin{tabular}{r@{\hspace{0.2cm}}     r@{\hspace{0.2cm}}rr@{\hspace{0.2cm}}r@     {\hspace{0.2cm}}rlcrrrrrl}
\hline
\multicolumn{6}{c}{Optical Position}&       \multicolumn{1}{c}{$z$} &       $N\sigma_c$    &\multicolumn{1}{c}{E}       & O$-$E &$\Delta r$      &\multicolumn{1}{c}{$S_{5}$}       & \multicolumn{1}{c}      {$\alpha^{5}_{1.4}$}      &Date \cr
\multicolumn{2}{c}{$\alpha$}&   \multicolumn{2}{c}{1950}&   \multicolumn{2}{c}{$\delta$}& &&mag &mag&  $''$ &mJy& &Obs.\cr\hline
00&03&22.31& 38&03&32.4&0.23 &1.75&17.10&$  2.43$& 0.97&   549&$   -0.07\phantom{^*}$&10/92            \cr                                      
00&20&45.98& 25&22&40.2&     &0.97&19.51&$> 2.68$& 2.72&   130&$  > 0.21\phantom{^*}$&10/92            \cr                                      
00&24& 2.78& 34&52& 6.6&0.33 &2.19&17.45&$  1.47$& 0.33&   453&$   -0.47\phantom{^*}$&10/92            \cr                                      
01&09&24.75& 35&06&24.9&0.45 &0.19&17.84&$  1.17$& 0.95&   362&$    0.00\phantom{^*}$&10/92            \cr                                      
01&19&58.40& 29&38&32.9&     &1.33&17.64&$  1.01$& 0.28&   154&$   -0.06\phantom{^*}$&10/92            \cr                                      
01&22& 9.46& 27&49&34.6&0.71 &1.87&18.19&$> 3.84$& 0.54&   228&$    0.25\phantom{^*}$&10/92            \cr                                      
01&29&46.04& 43&10& 8.7&     &0.87&18.67&$  1.16$& 0.89&   347&$    0.37\phantom{^*}$&10/92            \cr                                      
01&30&48.60& 38&07&36.8&     &0.24&19.02&$> 3.20$& 0.93&   204&$   -0.61\phantom{^*}$&10/92            \cr                                      
01&48&18.00& 25&02&38.9&3.10 &0.75&19.79&$> 1.62$& 0.30&   237&$    0.28\phantom{^*}$&10/92            \cr                                      
01&48&37.17& 27&29&52.8&1.26 &1.11&18.84&$  1.23$& 0.95&   991&$    0.28\phantom{^*}$&08/93            \cr                                      
02&17&49.86& 32&27&23.3&1.62 &1.43&17.44&$  1.32$& 1.06&   618&$   -0.34\phantom{^*}$&10/92            \cr                                      
02&29&29.47& 23&04&44.2&3.42 &2.53&19.93&$> 1.47$& 0.28&   457&$    0.18\phantom{^*}$&10/92            \cr                                      
02&30&18.23& 34&29&44.0&0.50 &2.47&18.77&$  2.08$& 0.97&   223&$   -0.03\phantom{^*}$&10/92            \cr                                      
03&25&37.21& 31&28&52.1&0.46 &1.89&18.98&$  1.57$& 0.09&   245&$    0.45\phantom{^*}$&10/92            \cr                                      
04&03&24.78& 25&03&28.2&0.68*&0.23&19.10&$  1.06$& 0.76&   212&$  > 0.59\phantom{^*}$&10/92            \cr                                      
06&10&52.79& 51&03& 7.9&1.59*&1.60&18.64&$  1.01$& 1.80&   202&$    0.20\phantom{^*}$&04/93            \cr                                      
07&07& 2.72& 47&37& 9.9&     &0.20&13.00&$  1.17$& 2.69&   906&$   -0.06\phantom{^*}$&04/93            \cr                                      
07&47& 2.64& 27&07&59.7&1.54*&1.45&19.58&$> 2.05$& 0.11&   139&$   -0.58\phantom{^*}$&04/93            \cr                                      
07&49&21.57& 37&38&10.2&0.44 &1.17&19.98&$> 1.82$& 1.62&   208&$   -0.04\phantom{^*}$&04/93            \cr                                      
07&49&35.26& 42&39&17.7&3.59 &0.19&18.08&$  1.65$& 0.94&   461&$   -0.33\phantom{^*}$&04/93            \cr                                      
07&53&10.94& 51&59& 2.2&1.33*&2.85&18.10&$  1.33$& 1.33&   196&$    0.43\phantom{^*}$&04/93            \cr                                      
08&05&50.51& 53&50&15.2&     &2.24&19.77&$  1.59$& 1.05&   197&$    0.07\phantom{^*}$&04/93            \cr                                      
08&30&11.07& 60&30& 3.3&0.72 &1.60&19.22&$  1.44$& 1.67&   309&$    0.14\phantom{^*}$&04/93            \cr                                      
08&30&31.96& 42&34&19.0&     &0.67&16.99&$  1.44$& 0.47&   341&$   -0.38\phantom{^*}$&04/93            \cr                                      
08&59&12.84& 43&22& 4.3&2.41 &1.15&19.55&$  1.10$& 1.21&   324&$   -0.02\phantom{^*}$&04/93            \cr                                      
09&00&58.72& 42&50& 0.4&     &0.25&19.90&$  1.75$& 0.99&   718&$   -0.52\phantom{^*}$&04/93            \cr                                      
09&03&18.88& 25&49&36.9&0.48 &2.51&19.04&$  2.38$& 1.32&   272&$   -0.62\phantom{^*}$&04/92            \cr                                      
09&08&43.63& 34&01&36.9&     &1.75&18.62&$  1.56$& 1.10&   250&$   -0.67\phantom{^*}$&04/92            \cr                                      
09&11&34.07& 35&24&31.6&1.07*&2.04&19.47&$  1.30$& 1.79&   321&$    0.28\phantom{^*}$&04/93            \cr                                      
09&16&11.95& 86&25&16.9&     &0.56&18.07&$  1.13$& 1.23&   253&$   0.21^*$&04/93            \cr                                      
09&20&19.83& 41&38&20.4&     &1.07&19.57&$  1.93$& 0.92&   183&$    0.12\phantom{^*}$&04/93            \cr                                      
09&24&51.80& 73&17&12.3&     &1.93&19.35&$> 2.30$& 0.21&   206&$    0.17\phantom{^*}$&04/93            \cr                                      
09&27&52.76& 35&16&49.9&     &0.12&19.18&$  1.24$& 1.12&   383&$   -0.08\phantom{^*}$&04/93            \cr                                      
09&37&21.97& 26&17& 7.1&     &1.03&19.21&$  1.64$& 0.76&   326&$    0.04\phantom{^*}$&04/92            \cr                                      
09&38&54.56& 27&42&19.0&     &2.39&17.73&$  1.15$& 2.80&   246&$    0.07\phantom{^*}$&04/93            \cr                                      
09&55&22.28& 50&54&19.0&1.15*&0.75&18.74&$  1.13$& 0.83&   173&$   -0.23\phantom{^*}$&04/93            \cr                                      
10&00&52.32& 26&19&45.9&     &1.56&18.31&$  1.21$& 0.20&   274&$   -0.46\phantom{^*}$&04/93            \cr                                      
10&09&51.79& 40&53&54.8&     &1.27&19.46&$> 2.31$& 1.84&   209&$   -0.58\phantom{^*}$&04/92            \cr                                      
10&13&58.30& 61&31&27.2&2.80 &0.65&18.12&$  1.03$& 0.27&   631&$    0.44\phantom{^*}$&04/93            \cr                                      
10&13&59.33& 20&52&47.8&3.11 &0.76&18.67&$  1.39$& 1.45&   976&$    0.18\phantom{^*}$&04/93            \cr                                      
10&17&26.37& 43&36& 0.8&1.96*&1.04&18.27&$  1.11$& 1.80&   232&$    0.33\phantom{^*}$&04/93            \cr                                      
10&23&13.10& 74&43&43.7&     &0.65&18.72&$  1.45$& 0.33&   204&$    0.03\phantom{^*}$&04/93            \cr                                      
10&30&32.56& 61&06&36.0&     &1.44&19.32&$  1.48$& 1.05&   579&$   -0.22\phantom{^*}$&04/93            \cr                                      
10&31&55.92& 56&44&18.5&0.46 &2.08&19.92&$> 2.27$& 0.37&  1200&$   -0.30\phantom{^*}$&04/93            \cr                                      
10&40&25.20& 24&24&19.4&     &1.21&17.09&$  1.09$& 0.35&   364&$    0.09\phantom{^*}$&04/93            \cr                                      
\hline
\end{tabular}
\end{center}
\label{comptab}
\end{table*}

\begin{table*}
\contcaption{}
\begin{center}
\begin{tabular}{r@{\hspace{0.2cm}}     r@{\hspace{0.2cm}}rr@{\hspace{0.2cm}}r@     {\hspace{0.2cm}}rlcrrrrrl}
\hline
\multicolumn{6}{c}{Optical Position}&       \multicolumn{1}{c}{$z$} &       $N\sigma_c$    &\multicolumn{1}{c}{E}       & O$-$E &$\Delta r$      &\multicolumn{1}{c}{$S_{5}$}       & \multicolumn{1}{c}      {$\alpha^{5}_{1.4}$}      &Date \cr
\multicolumn{2}{c}{$\alpha$}&   \multicolumn{2}{c}{1950}&   \multicolumn{2}{c}{$\delta$}& &&mag &mag&    $''$ &mJy& &Obs.\cr\hline
10&56&58.43& 21&13&27.9&0.40 &0.80&18.35&$  1.44$& 0.24&   230&$    0.53\phantom{^*}$&04/93            \cr                                      
10&58&26.65& 20&36&52.7&1.66*&0.92&19.79&$  1.43$& 0.47&   159&$  > 0.36\phantom{^*}$&04/93            \cr                                      
10&59&59.91& 79&49& 3.0&     &2.48&19.95&$> 2.06$& 0.36&   278&$   -0.40\phantom{^*}$&04/93            \cr                                      
11&05&35.16& 43&47& 9.5&1.22*&1.60&19.52&$  1.87$& 0.74&   375&$    0.27\phantom{^*}$&04/93            \cr                                      
11&07&46.02& 48&34& 9.6&0.74 &2.60&19.24&$> 2.32$& 0.25&   256&$   -0.51\phantom{^*}$&04/93            \cr                                      
11&08& 5.59& 85&28&15.3&2.85 &1.40&19.14&$  1.49$& 0.80&   108&$  0.29^*$&04/93            \cr                                      
11&08&48.70& 33&09&14.0&     &0.87&19.45&$  1.09$& 0.52&   205&$   -0.03\phantom{^*}$&04/93            \cr                                      
11&21&53.23& 23&24&24.2&     &0.59&19.29&$  2.33$& 1.06&   149&$  > 0.31\phantom{^*}$&04/92            \cr                                      
11&25&23.01& 59&41&46.3&     &0.29&19.95&$> 2.30$& 1.32&   393&$    0.06\phantom{^*}$&04/93            \cr                                      
11&38&26.30& 64&26&42.3&     &0.16&19.41&$  1.89$& 1.32&   334&$    0.15\phantom{^*}$&04/92            \cr                                      
11&40&55.77& 66&50& 9.4&2.32 &0.24&18.33&$  1.22$& 1.11&   246&$    0.19\phantom{^*}$&04/93            \cr                                      
11&43& 0.32& 44&37& 1.4&0.30 &2.52&17.29&$> 4.44$& 0.49&   245&$   -0.46\phantom{^*}$&04/93            \cr                                      
11&44& 4.47& 54&13&22.3&2.19*&1.17&19.63&$  1.58$& 1.29&   484&$    0.14\phantom{^*}$&04/93            \cr                                      
11&47&39.69& 43&48&46.4&3.02 &1.71&19.34&$  1.00$& 1.00&   187&$   -0.02\phantom{^*}$&04/93            \cr                                      
11&51&19.00& 40&53&32.7&0.92*&0.52&19.67&$  1.52$& 1.55&   380&$   -0.48\phantom{^*}$&04/93            \cr                                      
12&00&58.33& 31&05&43.4&     &1.12&19.97&$> 1.94$& 1.32&   150&$   -0.13\phantom{^*}$&04/93            \cr                                      
12&26&27.31& 49&14&35.6&     &0.89&18.58&$  1.24$& 0.39&   326&$   -0.24\phantom{^*}$&04/93            \cr                                      
12&26&48.63& 63&51&35.4&     &2.60&18.70&$> 3.35$& 0.68&   308&$   -0.07\phantom{^*}$&04/92            \cr                                      
12&27&44.35& 25&34&40.8&     &1.77&16.34&$  1.05$& 0.46&   425&$   -0.34\phantom{^*}$&04/93            \cr                                      
12&32&39.28& 36&37&49.6&1.60 &0.20&19.06&$  1.26$& 1.04&   250&$    0.65\phantom{^*}$&04/93            \cr                                      
12&42&26.34& 41&04&29.7&0.81 &0.52&19.07&$  1.45$& 0.64&   707&$   -0.65\phantom{^*}$&04/93            \cr                                      
12&46& 7.27& 58&36&49.5&     &2.33&14.10&$  1.31$& 0.29&   414&$    0.43\phantom{^*}$&04/93            \cr                                      
12&50&16.34& 33&27&15.3&     &0.28&19.44&$> 2.33$& 0.44&   182&$   -0.10\phantom{^*}$&04/92            \cr                                      
12&50&58.19& 53&17&27.2&     &0.97&16.38&$  1.00$& 0.85&   396&$   -0.24\phantom{^*}$&04/93            \cr                                      
12&57&18.83& 51&57& 5.9&     &0.15&19.87&$> 1.55$& 0.06&   247&$   -0.11\phantom{^*}$&04/93            \cr                                      
13&00&50.66& 69&18&57.3&0.57 &2.52&19.61&$  1.64$& 1.44&   190&$    0.12\phantom{^*}$&04/93            \cr                                      
13&08&41.32& 47&09&47.4&     &1.07&19.07&$  1.53$& 0.76&   393&$    0.07\phantom{^*}$&04/93            \cr                                      
13&10&32.91& 48&44&23.6&     &1.32&19.98&$> 1.84$& 0.78&   224&$    0.13\phantom{^*}$&04/93            \cr                                      
13&20&41.06& 39&27&47.5&2.98 &0.99&17.06&$  1.16$& 0.76&   230&$    0.57\phantom{^*}$&04/93            \cr                                      
13&22&21.87& 47&58&55.9&2.26 &2.07&19.47&$  1.07$& 1.03&   237&$    0.18\phantom{^*}$&04/93            \cr                                      
13&24&37.12& 22&26&22.5&1.40*&1.36&17.18&$  1.38$& 0.32&   849&$    1.01\phantom{^*}$&04/93            \cr                                      
13&25&10.66& 43&41&59.0&2.08 &0.96&18.45&$  1.33$& 1.28&   533&$   -0.22\phantom{^*}$&04/93            \cr                                      
13&30&47.33& 27&40&40.7&     &1.51&18.69&$  1.50$& 0.24&   161&$   -0.27\phantom{^*}$&04/93            \cr                                      
13&38&11.09& 38&09&53.5&3.10 &0.08&17.85&$  1.68$& 0.60&   305&$    0.00\phantom{^*}$&04/92            \cr                                      
13&38&56.79& 28&31&12.0&1.31*&0.64&19.55&$  2.15$& 1.08&   172&$  > 0.43\phantom{^*}$&04/92            \cr                                      
13&57&41.94& 76&57&53.3&     &0.59&18.39&$  1.54$& 0.69&   844&$    0.34\phantom{^*}$&04/93            \cr                                      
14&00& 7.22& 58&50& 8.0&     &0.68&18.90&$  1.02$& 0.59&   220&$    0.04\phantom{^*}$&04/93            \cr                                      
14&17&45.13& 27&20& 9.4&     &0.35&18.96&$  1.24$& 0.32&   414&$   -0.93\phantom{^*}$&04/93            \cr                                      
14&32& 9.41& 42&16&21.7&     &0.00&17.80&$  1.12$& 0.76&   353&$    0.19\phantom{^*}$&04/93            \cr                                      
14&33& 3.68& 20&34&22.1&     &0.35&18.28&$  1.32$& 0.68&   215&$   -0.43\phantom{^*}$&04/92            \cr                                      
14&35&49.41& 30&15& 3.8&     &1.87&19.38&$  1.02$& 0.45&   132&$  > 0.22\phantom{^*}$&04/93            \cr                                      
14&36&36.23& 44&31& 6.5&2.10 &0.96&18.69&$> 2.89$& 0.39&   279&$    0.37\phantom{^*}$&06/93            \cr                                      
14&39&54.00& 32&47& 5.4&2.12 &0.76&18.77&$  1.01$& 0.69&   340&$   -0.25\phantom{^*}$&04/93            \cr                                      
14&51&43.01& 27&00&43.6&     &0.85&18.85&$  1.23$& 0.64&   354&$   -0.44\phantom{^*}$&04/93            \cr                                      
14&56&46.89& 37&32&16.7&     &0.67&18.22&$  1.70$& 0.77&   591&$    0.45\phantom{^*}$&04/92            \cr                                      
\hline
\end{tabular}
\end{center}
\end{table*}
\begin{table*}
\contcaption{}
\begin{center}
\begin{tabular}{r@{\hspace{0.2cm}}     r@{\hspace{0.2cm}}rr@{\hspace{0.2cm}}r@     {\hspace{0.2cm}}rlcrrrrrl}
\hline
\multicolumn{6}{c}{Optical Position}&       \multicolumn{1}{c}{$z$} &       $N\sigma_c$    &\multicolumn{1}{c}{E}       & O$-$E &$\Delta r$      &\multicolumn{1}{c}{$S_{5}$}       & \multicolumn{1}{c}      {$\alpha^{5}_{1.4}$}      &Date \cr
\multicolumn{2}{c}{$\alpha$}&   \multicolumn{2}{c}{1950}&   \multicolumn{2}{c}{$\delta$}& &&mag &mag&    $''$ &mJy& &Obs.\cr\hline
15&08&45.19& 57&14& 2.6&4.30 &1.45&18.88&$> 3.45$& 0.45&   282&$    0.50\phantom{^*}$&04/92            \cr                                      
15&20& 4.78& 43&47&19.4&2.18 &0.19&18.37&$> 3.43$& 1.64&   245&$    0.40\phantom{^*}$&04/93            \cr                                      
15&26&12.05& 67&01&16.1&3.02 &0.89&17.15&$  1.22$& 0.72&   417&$   -0.02\phantom{^*}$&04/92            \cr                                      
15&32&31.55& 48&33&38.3&     &1.09&17.55&$> 4.17$& 1.15&   217&$   -0.30\phantom{^*}$&04/93            \cr                                      
15&33&58.63& 31&36&22.8&1.87 &2.57&19.65&$  1.66$& 0.67&   137&$   -1.26\phantom{^*}$&04/93            \cr                                      
15&38&54.02& 61&23&31.7&     &0.65&19.16&$  1.35$& 1.77&   252&$   -0.53\phantom{^*}$&04/92            \cr                                      
15&45&53.06& 49&46&18.2&0.70 &1.64&19.63&$  1.96$& 1.99&   549&$   -0.42\phantom{^*}$&04/93            \cr                                      
16&03&54.09& 20&40& 9.9&     &2.15&18.55&$  2.62$& 0.43&   182&$    0.24\phantom{^*}$&04/92            \cr                                      
16&16&34.54& 36&39&14.9&     &0.87&18.54&$  1.54$& 0.42&   268&$   -0.50\phantom{^*}$&04/92            \cr                                      
16&23&27.65& 57&48& 4.4&     &1.85&17.32&$  1.34$& 0.71&   590&$    0.13\phantom{^*}$&04/92            \cr                                      
16&29&55.14& 49&34& 0.2&0.52 &1.36&18.31&$  1.86$& 0.30&   394&$    0.16\phantom{^*}$&04/93            \cr                                      
16&30&17.64& 74&37&22.1&0.70 &0.49&18.68&$  1.99$& 1.49&   175&$   -0.63\phantom{^*}$&04/93            \cr                                      
16&30&30.00& 77&13&15.4&     &2.07&18.58&$  1.07$& 0.26&   174&$   -0.83\phantom{^*}$&04/93            \cr                                      
16&46&50.83& 41&09&16.0&0.85 &1.48&19.42&$  1.00$& 0.92&   191&$   -0.51\phantom{^*}$&04/93            \cr                                      
16&51&16.73& 39&07&42.8&     &0.53&19.87&$  1.32$& 1.14&   308&$   -0.07\phantom{^*}$&04/93            \cr                                      
17&07&56.14& 77&59&52.1&     &1.05&18.16&$  1.00$& 0.71&   232&$   -0.10\phantom{^*}$&04/93            \cr                                      
17&14& 3.75& 21&55&29.4&     &0.75&19.27&$  1.78$& 0.80&   716&$    0.07\phantom{^*}$&04/92            \cr                                      
17&18& 5.57& 23&38&28.8&1.86*&0.80&19.09&$> 2.94$& 0.09&   150&$   -0.43\phantom{^*}$&04/93            \cr                                      
17&45&47.98& 62&27&55.2&3.89 &0.03&18.29&$  2.61$& 1.06&   580&$   -0.22\phantom{^*}$&04/92            \cr                                      
17&47&29.53& 43&22&42.4&     &0.64&16.93&$  1.00$& 0.80&   367&$    0.06\phantom{^*}$&04/93            \cr                                      
17&50&20.94& 50&56&16.6&     &2.93&18.86&$> 2.92$& 1.17&   192&$   -0.38\phantom{^*}$&04/92            \cr                                      
17&59&33.84& 75&39&22.4&3.05 &0.31&16.12&$  1.16$& 0.50&   145&$  > 0.29\phantom{^*}$&04/93            \cr                                      
21&57&53.68& 21&23&31.3&     &0.67&18.24&$  1.06$& 0.19&   251&$    0.05\phantom{^*}$&10/92            \cr                                      
22&14&52.80& 20&09&47.5&     &0.97&16.43&$  1.24$& 1.32&   221&$   -0.48\phantom{^*}$&06/94            \cr                                      
22&23&13.32& 20&24&58.5&3.56 &0.93&18.38&$  1.76$& 0.86&   143&$   -0.31\phantom{^*}$&06/94            \cr                                      
22&46&34.80& 20&51& 9.5&     &1.79&18.92&$  1.03$& 0.94&   771&$    0.12\phantom{^*}$&10/92            \cr                                      
23&07& 9.24& 26&52&15.5&0.27 &2.65&18.78&$  3.10$& 1.21&   137&$   -0.54\phantom{^*}$&10/92            \cr                                      
23&19&57.66& 44&29&13.9&     &0.35&19.92&$  1.41$& 0.95&   366&$    0.14\phantom{^*}$&10/92            \cr                                      
23&41&51.56& 29&35&41.4&     &1.19&18.75&$  1.15$& 0.45&   190&$   -0.35\phantom{^*}$&10/92            \cr                                      
23&42&20.71& 34&17& 9.3&2.99 &1.17&18.33&$  1.19$& 0.68&   146&$    0.07\phantom{^*}$&10/92            \cr                                      
\hline
\end{tabular}
\end{center}
\end{table*}

\begin{table*}
\caption{Objects observed which do not satisfy the selection criteria of the complete sample.}
\begin{center}
\begin{tabular}{r@{\hspace{0.2cm}}     r@{\hspace{0.2cm}}rr@{\hspace{0.2cm}}r@     {\hspace{0.2cm}}rlcrrrrrl}
\hline
\multicolumn{6}{c}{Optical Position}&       \multicolumn{1}{c}{$z$} &       $N\sigma_c$    &\multicolumn{1}{c}{E}       & O$-$E &$\Delta r$      &\multicolumn{1}{c}{$S_{5}$}       & \multicolumn{1}{c}      {$\alpha^{5}_{1.4}$}      &Date \cr
\multicolumn{2}{c}{$\alpha$}&   \multicolumn{2}{c}{1950}&   \multicolumn{2}{c}{$\delta$}& &&mag &mag&     $''$ &mJy& &Obs.\cr\hline
00&35&41.55& 41&20&37.2&1.35*&3.87&19.36&$  1.36$& 0.31&  1114&$    0.73$&10/92            \cr                                      
08&24&10.19& 52&27&54.0&0.34 &3.13&18.10&$  2.16$& 0.86&   275&$   -0.27$&04/93            \cr                                      
09&02& 5.61& 28&00&18.1&0.86*&4.53&18.66&$  1.22$& 1.31&   168&$   -0.26$&04/93            \cr                                      
09&58&17.68& 29&26& 4.3&     &0.76&18.22&$  0.89$& 0.47&   162&$  > 0.38$&04/92            \cr                                      
10&59&39.86& 59&57&28.5&1.83 &1.24&16.73&$  0.76$& 0.17&   250&$   -0.21$&04/93            \cr                                      
11&00&29.73& 30&30&52.3&0.38 &1.76&17.50&$  0.53$& 0.46&   211&$   -0.14$&04/93            \cr                                      
11&24&13.75& 45&32&36.6&1.81 &0.23&16.98&$  0.41$& 0.46&   355&$   -0.26$&04/93            \cr                                      
11&28& 3.36& 30&48& 8.0&0.74 &0.99&17.73&$  0.70$& 0.76&   377&$   -0.16$&04/93            \cr                                      
11&45&24.06& 26&52&22.0&0.87 &0.63&17.40&$  0.54$& 0.71&   381&$   -0.07$&04/93            \cr                                      
12&02& 4.32& 52&45&23.5&2.73 &0.32&17.88&$  0.67$& 0.14&   228&$   -0.52$&04/93            \cr                                      
12&06&51.10& 41&36&22.4&     &0.48&16.35&$  0.86$& 0.48&   515&$    0.37$&04/93            \cr                                      
12&18&59.03& 44&28& 7.6&1.35 &1.28&17.35&$  0.73$& 0.27&   478&$   -0.18$&04/93            \cr                                      
15&50&49.07& 78&54&23.2&     &3.03&19.02&$> 2.96$& 2.76&   166&$   -0.77$&04/93            \cr                                      
15&57&41.39& 56&33&41.4&     &0.40&16.65&$  0.99$& 0.91&   215&$   -0.38$&04/92            \cr                                      
17&53&56.18& 64&52&29.0&     &3.60&18.30&$  1.52$& 0.27&   267&$    0.61$&04/92            \cr                                      
22&09&45.67& 23&40&49.2&     &0.61&18.89&$  0.70$& 0.62&  1212&$    0.69$&10/92            \cr                                      
22&26&12.04& 27&51&40.1&0.59*&1.65&18.70&$  0.14$& 0.41&   160&$   -0.53$&10/92            \cr                                      
23&21&14.36& 27&18&50.8&1.69*&0.53&18.95&$  0.96$& 1.35&   186&$   -0.38$&10/92            \cr                                      
\hline
\end{tabular}
\end{center}
\label{extratab}
\end{table*}

\begin{table*}
\caption{Objects from the complete sample with no spectra or very poor spectra.}
\begin{center}
\begin{tabular}{r@{\hspace{0.2cm}}     r@{\hspace{0.2cm}}rr@{\hspace{0.2cm}}r@     {\hspace{0.2cm}}rcrrrrrl}
\hline
\multicolumn{6}{c}{Optical Position}&       $N\sigma_c$    &\multicolumn{1}{c}{E}       & O$-$E &$\Delta r$      &\multicolumn{1}{c}{$S_{5}$}       & \multicolumn{1}{c}      {$\alpha^{5}_{1.4}$}      &Comments \cr
\multicolumn{2}{c}{$\alpha$}&   \multicolumn{2}{c}{1950}&   \multicolumn{2}{c}{$\delta$}&&mag &mag&  \multicolumn{1}{c}{$''$} &mJy& & \cr\hline
 00&55&47.54& 32&55& 7.1&0.53&19.87&$  1.13$& 1.10&   164&$   -0.15$&WHT 10/92 No object detected\cr                                      
 03&58&49.05& 21&02& 8.3&0.09&17.94&$  1.23$& 0.43&   158&$   -0.68$&No spectrum taken      \cr                                      
 13&21&58.84& 41&03&48.9&0.16&19.49&$  2.56$& 0.57&   413&$    0.11$&WHT 04/92 low s/n \cr                                      
 16&56&24.93& 48&13& 3.8&0.04&19.55&$> 2.37$& 0.83&   847&$   -0.12$&WHT 04/93 No object detected\cr                                      
\hline
\end{tabular}
\end{center}
\label{badspectab}
\end{table*}

After the 1993 August observing run the spectroscopic observations of
the complete sample defined in Section 1 were essentially complete to
the limit of the POSS-I plates, $\rm E\sim 20$.  The spectra of
objects that satisfy all the selection criteria are shown in
Fig.~\ref{jbspec}, and their properties are tabulated in
Table~\ref{comptab}. Spectra of the extra objects are shown in
Fig.~\ref{jbothers}, and their properties given in
Table~\ref{extratab}. Some of the ISIS spectra of lower
signal-to-noise ratio have been smoothed (using a Gaussian smoothing
function with $\sigma=2$ pixels). Three objects in the complete sample
still have inconclusive spectra. In two cases the objects were not
detected (possibly because the POSS-I identifications were chance
coincidences with stars that have since moved), and in the third case
the signal-to-noise ratio was too low.  One other object remains to be
observed.  The objects with poor or no spectra are listed in
Table~\ref{badspectab}. Therefore, of the 123 objects requiring
spectroscopy, we now present spectra for 119, plus one previously
known object that was re-observed (GB0916+8625).

In the complete sample a total of 11 new quasars with $z>3$ were
found, of which four have $z>3.5$. Since 120 objects were observed, the
efficiency of the survey for $z>3$ QSOs is approximately 1 in 10.  The
$z>3$ QSOs are marked on the colour-magnitude diagram
(Fig.~\ref{jbcolmag}) with star symbols.  A further 12 new quasars in the
range $2<z<3$ were found (but note that the colour selection criterion
means that the sample is substantially incomplete for $z<3$).  The highest
redshift object found was a quasar with a redshift of $z=4.30$, the
first known radio-selected quasar with $z>4$. This object has been the
subject of an earlier paper (Hook et al. 1995).

\begin{figure}
\centerline{
\psfig{figure=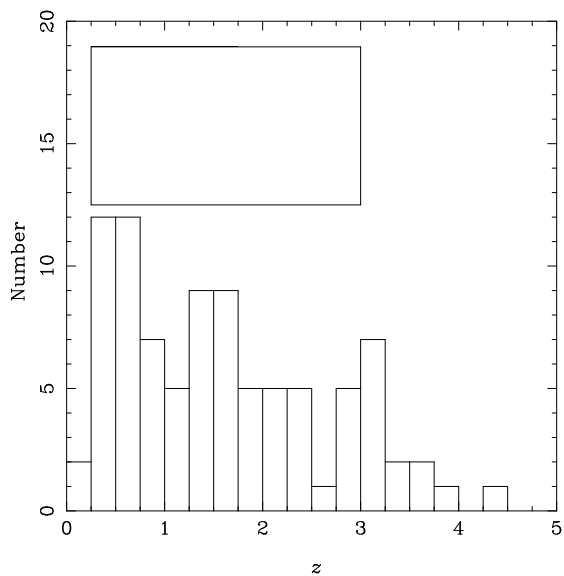,height=3.0in,bbllx=39pt,bblly=123pt,bburx=525pt,bbury=624pt}
}
\caption[]{Histogram of redshifts obtained for the complete sample.
The histogram includes the previously known objects and objects with
uncertain redshifts, giving a total of 90 objects. Objects with
spectra from which redshifts could not be measured (71 of the complete
sample) are represented by the rectangle. It is unlikely that any of
these have $z>3$ (see text for explanation).}
\label{jb_nz}
\end{figure}

The remaining objects were found to be low-redshift ($z<2$) quasars,
emission-line galaxies and objects with featureless spectra which
could be ruled out as possible high-redshift quasars.  Redshift
estimates are now available for 90 objects in the complete sample
(including previously known objects), and their redshift distribution
is shown in Fig.~\ref{jb_nz}.

Assuming the background source density of red, stellar objects to be
0.002$\rm arcsec{^-2}$, we expect a false identification rate in a 3.0
arcsec radius of 0.057 (see Section 3), or nine objects in the sample
of 161. This compares to the two objects that were not detected and
could be chance coincidences with stars that have since moved. It is
also possible that some of the objects that we found to have `featureless'
spectra are chance coincidences with foreground stars.

The $z>3$ QSOs will be combined with previously existing data to
calculate the luminosity function, the subject of a future paper.

\subsection{The new sample of $z>3$ QSOs}

\begin{figure}
\centerline{
\psfig{figure=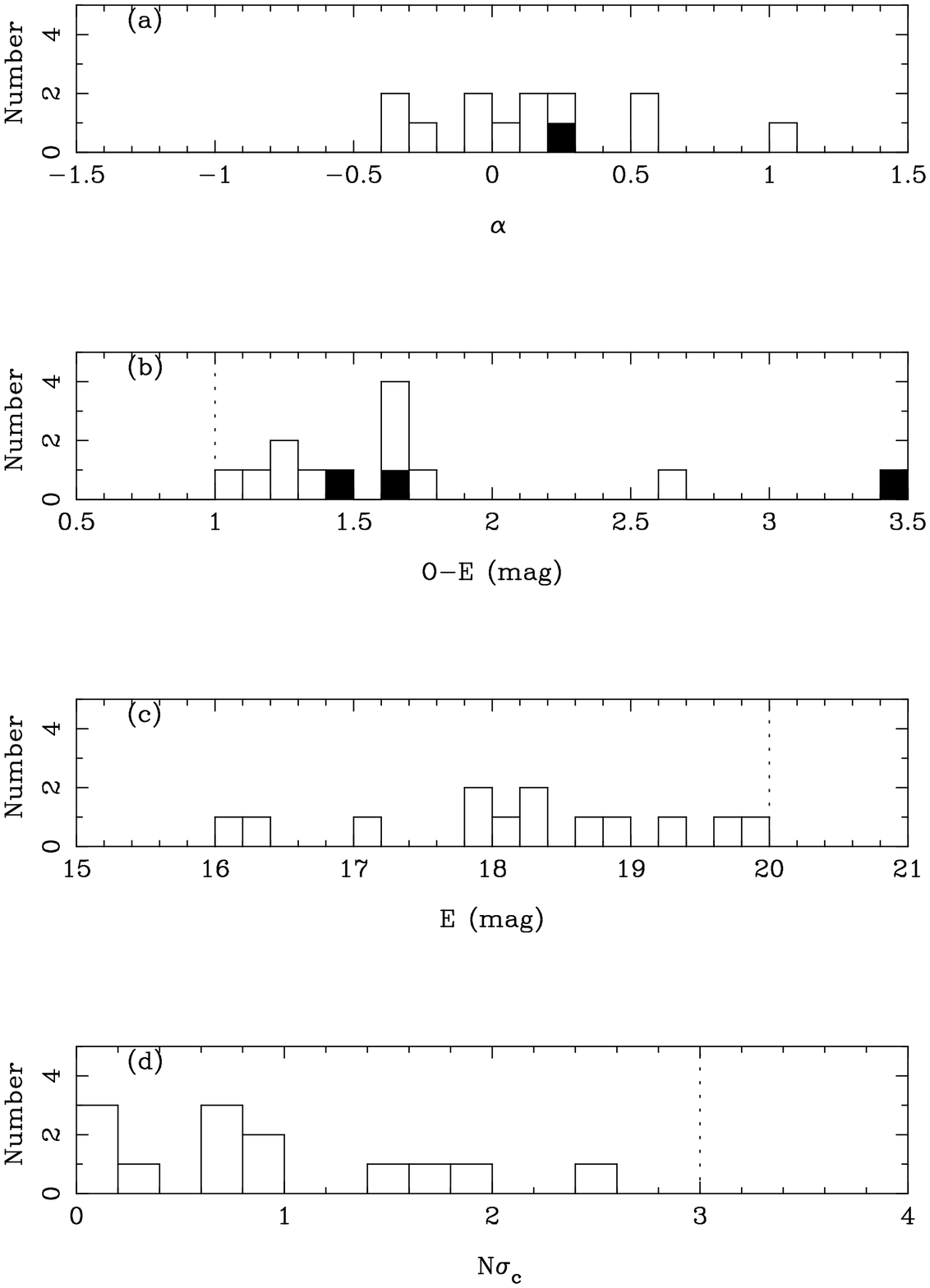,height=4.7in,bbllx=35pt,bblly=18pt,bburx=547pt,bbury=725pt}
}
\caption[]{The distributions of the 13 $z>3$ quasars of the complete sample
in terms of (a) spectral index $\alpha$, (b) colour $\rm O-E$, (c) apparent E
magnitude and (d) classification parameter $N\sigma_c$. The filled
histogram in (b) indicates the contribution in each bin from objects
that did not appear on the O plate, hence these $\rm O-E$ values are
lower limits. Similarly in (a) the filled histogram shows the object
with a lower limit on $\alpha$. The vertical dotted lines show the limits
imposed on the parameters in constructing the sample.}
\label{z3hists}
\end{figure}

\begin{figure*}
\centerline{
\psfig{figure=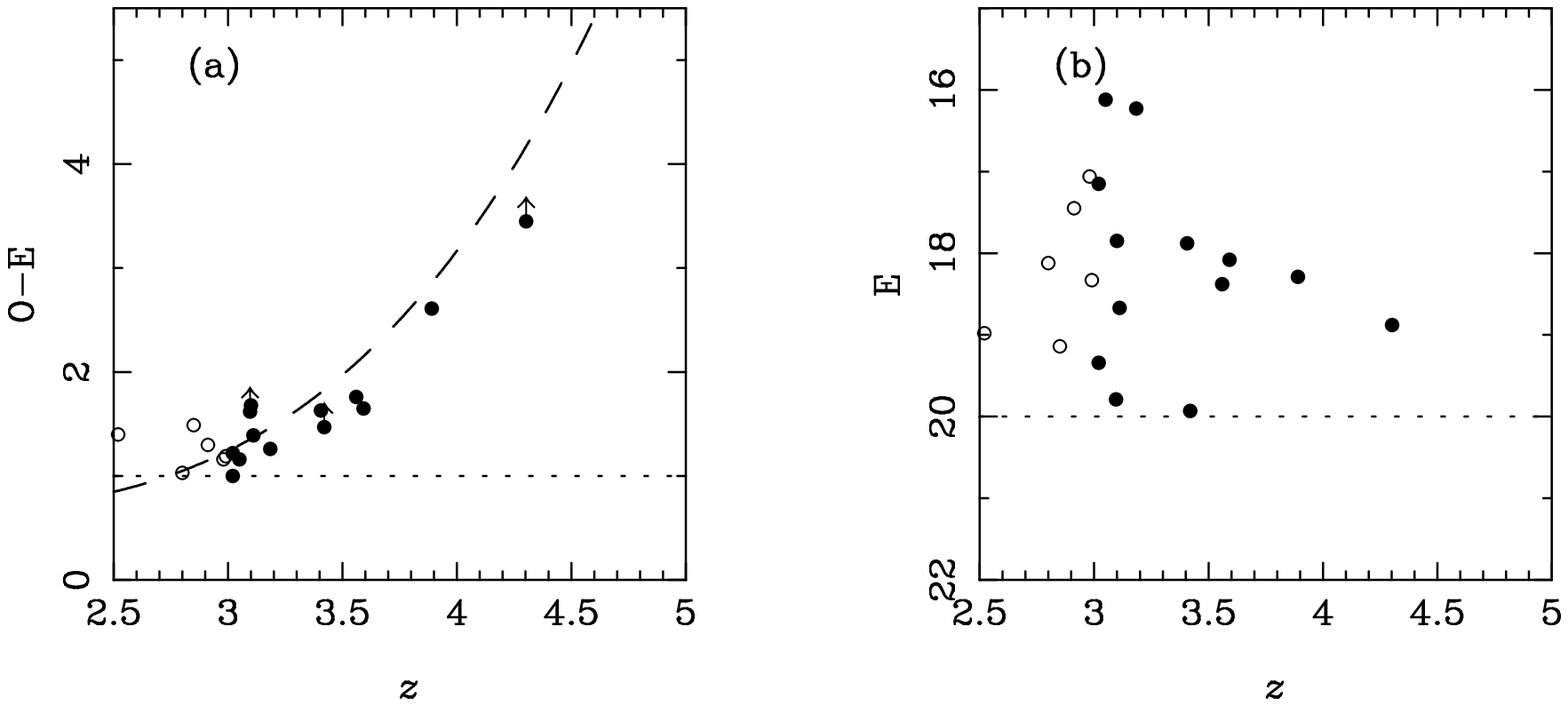,height=2.5in,bbllx=33pt,bblly=260pt,bburx=547pt,bbury=500pt}
}
\centerline{
\psfig{figure=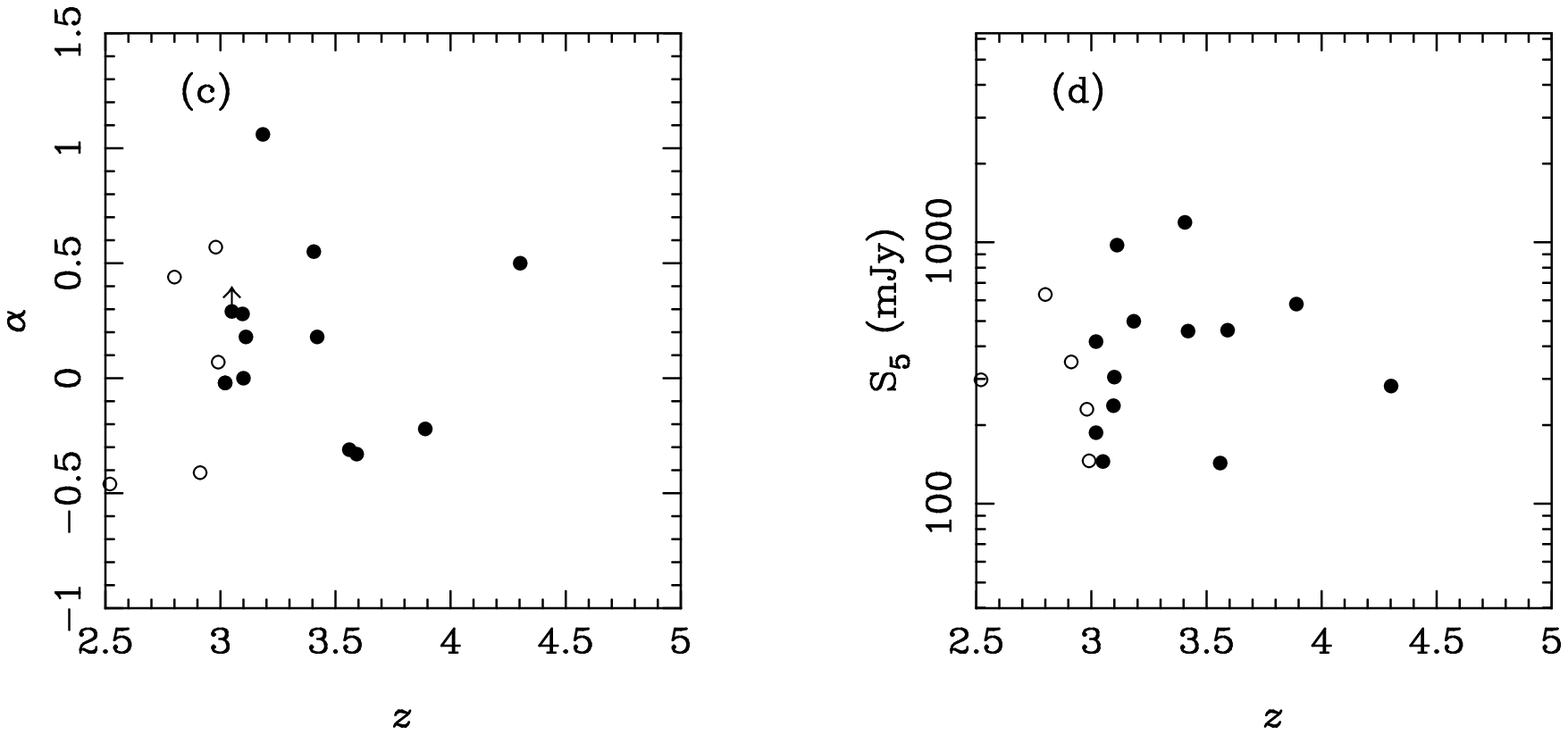,height=2.5in,bbllx=33pt,bblly=260pt,bburx=547pt,bbury=500pt}
}
\caption[]{(a) $\rm O-E$ colour versus {\it z}, (b) E magnitude versus {\it z},
(c) spectral index $\alpha$ versus {\it z}, (d) 5-GHz Radio flux $S_5$
versus {\it z} for objects with $z>2.5$ in the complete sample.  The
horizontal dotted lines in (a) and (b) show the limits imposed on the
parameters in constructing the sample. The dashed curve in (a) shows a
fit obtained using $\rm O-E$ versus $z$ data for a sample of optically
selected quasars.  Objects that did not appear on the O plate are
shown with arrows in (a).  Objects with $2.5<z<3.0$ are shown with
open circles -- the survey is substantially incomplete for these
objects.}
\label{z3plots}
\end{figure*}

A summary of the 13 $z>3$ QSOs in the complete sample is given in
Table~\ref{z3tab}.  Histograms showing the distributions of spectral
index, colour, E magnitude and classification parameter $N\sigma_c$
for all the quasars with $z>3$ included in the new sample are shown in
Fig.~\ref{z3hists}.  The distributions of colour, E magnitude,
spectral index and 5-GHz flux $S_5$ with redshift are shown in
Fig.~\ref{z3plots}.  As can be seen from these, all the $z>3$
quasars lie well within the criterion for having a stellar
(unresolved) optical image ($N\sigma_c\le 3.0$).  In addition, most of
the quasars with $z>3$ have O$-$E colours significantly greater than
the imposed limit of 1.0.  This suggests that any incompleteness
introduced by the colour cut and classification criterion is small.
The distribution of spectral indices appears to have a mean of about
zero, and no obvious trend with redshift is seen.  Incompleteness
introduced by other factors, such as the optical limit for
identifications, will be discussed in a future paper, although it is
interesting to note that the distribution of E magnitudes is fairly
flat, and does not `pile up' at the limit of $\rm E=20$.

\begin{table}
\caption{$z>3$ objects in the complete sample. See Table 1 for an explanation of the column headings.}
\begin{center}
\begin{tabular}{llcrrrr}
\hline
{Name}&       \multicolumn{1}{c}{$z$} &\multicolumn{1}{c}{E}       & O$-$E &\multicolumn{1}{c}{$S_{5}$}       & \multicolumn{1}{c}      {$\alpha^{5}_{1.4}$}\cr
 & &mag &mag&mJy\cr\hline
GB0148+2502&3.10 &19.79&$> 1.62$&   237&$    0.28$\cr
GB0229+2304&3.42 &19.93&$> 1.47$&   457&$    0.18$\cr
GB0636+6801&3.18 &16.23&$  1.26$&   499&$    1.06$\cr
GB0642+4454&3.41 &17.88&$  1.63$&  1191&$    0.55$\cr
GB0749+4239&3.59 &18.08&$  1.65$&   461&$   -0.33$\cr
GB1013+2052&3.11 &18.67&$  1.39$&   976&$    0.18$\cr
GB1147+4348&3.02 &19.34&$  1.00$&   187&$   -0.02$\cr
GB1338+3809&3.10 &17.85&$  1.68$&   305&$    0.00$\cr
GB1508+5714&4.30 &18.88&$> 3.45$&   282&$    0.50$\cr
GB1526+6701&3.02 &17.15&$  1.22$&   417&$   -0.02$\cr
GB1745+6227&3.89 &18.29&$  2.61$&   580&$   -0.22$\cr
GB1759+7539&3.05 &16.12&$  1.16$&   145&$  > 0.29$\cr
GB2223+2024&3.56 &18.38&$  1.76$&   143&$   -0.31$\cr
\hline
\end{tabular}
\end{center}
\label{z3tab}
\end{table}

Five of the quasars in Table~\ref{z3tab} are radio sources included in
the CJ1 (Polatidis et al. 1995) or CJ2 survey (Taylor et al.
1994). These are GB0636+6801, GB0642+4454, GB0749+4239, GB1526+6701
and GB1745+6227.

Two $z>3$ quasars are known to lie within the survey area but were not
found by this survey. A gravitational lens, 1422+2309, with a redshift
of 3.62 has been discovered as part of the new VLA/MERLIN search for
small-separation lenses (Patnaik et al. 1992b).  This source is
included in the 0.2Jy sample and satisfies all the selection criteria
used to define the complete sample of $z>3$ quasars except that its
optical image is classified as `merged' by the APM software, and it
did not meet the $N\sigma_c\le 3.0$ criterion. It was therefore not
included in the spectroscopic sample. Also a $z=3.1$ quasar,
GB1839+389, was found as the result of spectroscopic observations of
CJ2 sources (Henstock et al., in preparation). This object is also
included in the 0.2Jy sample but the optical image was classified as
highly non-stellar on the E plate.

Finding charts for the 13 $z>3$ quasars in the complete sample are
given in Fig.~\ref{fct}.

\subsection{Notes on individual objects}

\noindent{\it GB0003+3803.} Stickel \& K\"uhr (1994) also find $z=0.23$. 

\noindent{\it GB0024+3452.} Stickel \& K\"uhr (1993b) also find $z=0.33$.

\noindent{\it GB0035+4120.} Stickel \& K\"uhr (1993b) also find $z=1.35$.

\noindent{\it GB0707+4737.} Stickel \& K\"uhr (1994) give the 
identification as an 18.2-mag quasar with $z=1.292$. However, the APM
scan and the digitized sky survey image show only one bright ($\rm
E\sim 13$) object at the radio position. Visual inspection of the
POSS-I E and O plates and the POSS-II $B_{\rm J}$ plate of the same
field reveals that GB0707+4737 is slightly elliptical on the POSS-II
plate and there is evidence for a second object in the south-easterly
direction on the POSS-I O plate. A contour plot given in Meisenheimer
\& R\"oser (1983) shows the fainter identification close to the bright
star.

\noindent{\it GB0927+3516.} Henstock (in preparation) also finds
this to be featureless after 2000s on the Isaac Newton Telescope.

\noindent{\it GB1030+6106.} A tentative redshift $z=0.336$ is given by Stickel \& K\"uhr (1994).

\noindent{\it GB1031+5644.} Aller, Aller \& Hughes (1992) report $z=0.45$ (compared with our measurement of $z=0.46$).

\noindent{\it GB1105+4347.} Henstock (in preparation) finds $z=1.226$
also.

\noindent{\it GB1124+455.} Henstock (in preparation) also finds $z=1.811$, as do Bade et al. (1995).

\noindent{\it GB1144+5413.} Stickel \& K\"uhr (1994) find $z=2.201$ and  Xu et al. (1994) find $z=2.20$ (compared with our tentative $z=2.19$).

\noindent{\it GB1145+2652.} Bade et al. (1995) also find $z=0.867$.

\noindent{\it GB1151+4053} Henstock (in preparation) finds $z=0.916$.
Our spectrum is consistent with this.

\noindent{\it GB1202+5245} Owen, Ledlow \& Keel (1995) also find $z=2.73$.

\noindent{\it GB1206+4136} Henstock (in preparation) also finds this to be 
featureless after 2500s on the INT.

\noindent{\it GB1218+4428} Henstock (in preparation) also finds $z=1.345$.

\noindent{\it GB1242+4104} Xu et al. (1994) also find $z=0.813$.

\noindent{\it GB1246+5836} This is classified as a featureless BL Lac by 
Vermeulen \& Taylor (1995) and by Fleming et al. (1993).

\noindent{\it GB1250+5317} Henstock (in preparation) also finds this 
to be featureless after 600s on the INT.

\noindent{\it GB1325+4341} Stickel \& K\"uhr (1994) give $z=2.073$ (compared 
with our $z=2.08$).

\noindent{\it GB1432+4216} Vermeulen et al. (1996) 
find $z=1.240\pm 0.004$ for this source.

\noindent{\it GB1456+3732} Vermeulen et al. (1996) obtain
$z=0.333\pm 0.001$ for this source, based on weak [OII], [OIII] and
H$\alpha$ lines that are not visible in our spectrum.

\noindent{\it GB1508+5714}
This is the highest redshift object found in our sample, a quasar with
$z=4.30$. This quasar is the first known radio-selected quasar with
$z>4$.  It has an O$-$E colour of $>3.44$ and an E magnitude of 18.9,
and is such a red object that it did not appear on the O plate.

\noindent{\it GB1623+5748} Vermeulen et al. (1996) 
obtain a redshift $z=0.789\pm 0.001$ for this source based on the
[OIII] doublet, not visible in our spectrum.

\noindent{\it GB1745+6227}
This object has been independently discovered by Becker, Helfand \&
White (1992) on the basis of its X-ray emission. They report a
redshift of 3.87, which is lower than the value derived here. Stickel
\& K\"uhr (1993a) report a redshift of 3.886, in close agreement with
the value derived in this work.

\noindent{\it GB1747+4322} Vermeulen et al. (1996) also find this 
to be featureless after 3000s on the Hale telescope.

\noindent{\it GB1759+7539} This object, with $z=3.05$, is optically very bright
(E=16.1 mag).

\noindent{\it GB2319+4429} Henstock (in preparation) 
also finds this to be featureless after 3000s on the INT.

\noindent{\it GB2342+3417} White, Kinney \& Becker (1993) report $z=3.01$
compared with our measurement of $z=2.99$.

\begin{figure*}
\centerline{
\psfig{figure=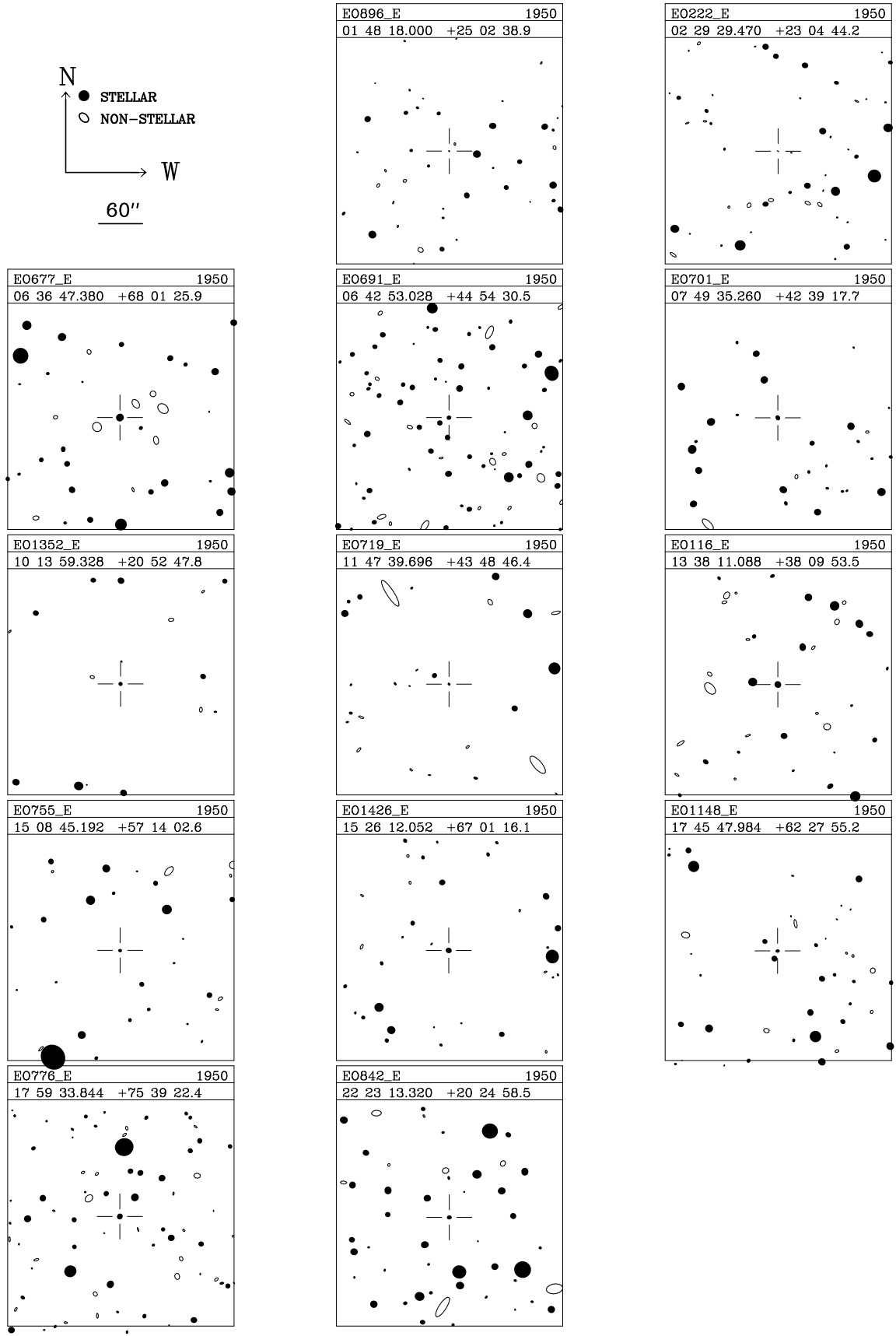,height=9.0in,bbllx=48pt,bblly=14pt,bburx=583pt,bbury=778pt}
}
\caption[]{Finding charts from APM scans of POSS-I E
plates ($\rm \sim R$ band) for the $z>3$ QSOs in the complete sample.}
\label{fct}
\end{figure*}

Finding charts and spectra for GB1508+5714 and GB1745+6227 have
already been given by Hook et al. (1995), and are included here for
completeness (Figures~\ref{jbspec} and~\ref{fct}).

\section{Conclusions}
In conclusion, we have shown that the radio/colour selection method is
a highly successful technique for finding $z>3$ QSOs. This paper
provides the first large, well-defined sample of radio-loud QSOs with
$z>3$.

We are obtaining follow-up observations of two other radio samples,
the 100-mJy `MG-VLA' sample (Lawrence et al. 1986) and a 50-mJy sample
(Leh\'ar 1991).  These samples combined with that used in this paper
contain a total of $\sim 6000$ radio sources, and the total area
optically identified using the APM covers 3.7sr (12 000 $\rm deg^2$)
of sky.

\section*{ACKNOWLEDGMENTS}
We are very grateful to Alok Patnaik, Ian Browne, Peter Wilkinson and
Joan Wrobel for supplying the radio catalogue used in this work prior
to publication. We thank Lisa Storrie-Lombardi for help in obtaining
the spectra and reducing the data for GB1745+6227. We also thank Rene
Vermeulen for checking our redshifts against those in the CJ1/CJ2 data
base.  IMH acknowledges a NATO postdoctoral fellowship and RGM thanks
the Royal Society for support.  This research has made use of the
NASA/IPAC Extragalactic Database (NED) which is operated by the Jet
Propulsion Laboratory, Caltech, under contract with the National
Aeronautics and Space Administration. IRAF is distributed by the
National Optical Astronomy Observatories, which is operated by the
Association of Universities for Research in Astronomy, Inc. (AURA)
under cooperative agreement with the National Science Foundation.

\end{document}